\documentstyle[twocolumn,aps,prd,epsfig,times,floats]{revtex}
\draft

\begin{document}

\title{Propagation of ultra-high energy protons in the nearby universe}

\author {Todor Stanev$^1$, Ralph Engel$^1$, Anita M\"{u}cke$^{2,3}$,
 Raymond J. Protheroe$^2$ \& J\"{o}rg P.~Rachen$^4$}
\address {
$^1$Bartol Research Institute, 
University of Delaware, Newark, DE 19716, USA
}
\address {$^2$ Dept. of Physics \& Math Physics, University of Adelaide,
SA 5005, Australia
}
\address {$^3$ Universit\'e de Montr\'eal, D\'epartement de Physique,
 Montr\'eal, H3C3J7, Canada
}
\address {$^4$ Sterrenkundig Instituut, Universiteit Utrecht,
 3584 CC Utrecht, The Netherlands
}
%\widetext

\date{Submitted to Phys.~Rev.~D: 31 March 2000; accepted 21 July 2000}

\wideabs{
\maketitle
\begin{abstract}

 We present a new calculation of the propagation of protons with energies
 above $10^{19}$ eV over distances of up to several hundred Mpc. The
 calculation is based on a Monte Carlo approach using the event generator
 SOPHIA for the simulation of hadronic nucleon-photon interactions  and a
 realistic integration of the particle trajectories in a random extragalactic
 magnetic field. Accounting for the proton scattering in the magnetic field
 affects noticeably the nucleon energy as a function of the distance
 to their source and allows us to give realistic predictions on arrival
 energy, time delay, and arrival angle distributions and correlations as
 well as secondary particle production spectra.
\end{abstract}
%\date{Submitted to Phys Rev D, ???? 2000}
}

% \narrowtext

\flushbottom

\section{Introduction}

  The world statistics of ultra high energy cosmic ray (UHECR) events
 of energy above 10$^{20}$ eV has now grown to 20
 events~\cite{SLC,AGASA}.  It is very difficult to accelerate
 particles to such high energies in astrophysical shocks, the process
 thought to be responsible for the majority of the galactic cosmic
 rays~\cite{Hillas84}. This has led to a large number of production
 models, many of them based on exotic particle physics
 scenarios~\cite{BS00}. The gyroradii of 10$^{20}$ eV protons are
 significantly larger than our own Galaxy and this suggests an
 extragalactic origin~\cite{Cocconi} for any astrophysical scenario
 ($r_g = 100 {\rm kpc} \times (E/10^{20} {\rm eV}) \times (1 \mu{\rm
 G}/B)$ with $E$ and $B$ being the proton energy and the magnetic
 field strength, respectively).  The large distances between potential
 UHECR sources and Earth leads to another set of problems first
 pointed out independently by Greisen and by Zatsepin \& Kuzmin, now
 widely known as the GZK effect~\cite{GZK}. UHECR protons interact
 with photons of the microwave background radiation and lose their
 energy relatively rapidly during propagation over distances of tens
 of megaparsecs. This should result in a cutoff in the cosmic ray
 spectrum at an energy just below 10$^{20}$ eV.

 Many different
 calculations~\cite{Hill85,BG88,YT93,RB93,AhCronin,PJ96,Lee98},
 performed using various techniques, of the modification of the cosmic
 ray spectrum due to propagation have been published since the
 original suggestion.  As a result, the general features of the cosmic ray
 spectrum after propagation are well established.  Differences between
 the various approaches are, however, significant and the accuracy
 achieved is not sufficient for the interpretation of the existing
 experimental data, and more accurate calculations are needed for the
 expected significant increase of the experimental
 statistics~\cite{HiRes,Auger,TelArr,OWL}.

 Previous calculations can be divided into two classes dealing mainly
 with: (a) the energy loss
 processes~\cite{Hill85,BG88,YT93,RB93,AhCronin,PJ96,Lee98}, and (b)
 the deflection and scattering of protons in the extragalactic
 magnetic field~\cite{Lampard97,Clay98a,Acht99}. The first group of
 calculations shows that small differences in the realization of the
 proton energy loss processes generate observable differences in the
 predicted spectra at Earth. Such calculations, however, cannot
 establish an accurate relation between the distance of a potential
 source and the modification of the proton spectrum emitted by this
 source because the influence of the extragalactic magnetic field is
 neglected. 
Among the calculations of the second kind,
Refs.~\cite{WME96,Lampard97,Clay98a} do not consider the proton
energy losses in a satisfactory way, and
%Refs.~\cite{Medina97,Sigl97a,Sigl99a,Acht99} mostly discuss their
Refs.~\cite{Medina97,Sigl97a,Sigl99a} mostly discuss their
results in a specific context. Only Achterberg {\em et al.}
\cite{Acht99,Achterror} give a detailed discussion of the fundamental
aspects of UHECR propagation in extragalactic magnetic fields, which we
are interested in here.

 We present here calculations performed with the photoproduction event
 generator SOPHIA~\cite{SOPHIA}, which is proven to reproduce well the
 cross section and final state composition in nucleon-photon
 interactions for energies from the particle production threshold up
 to hundreds of GeV in the center-of-mass system.  We also account for
 all other energy loss processes of UHECR nucleons, and calculate the
 proton deflection in the extragalactic magnetic field in three
 dimensions.

 We restrict ourselves to proton injection energies up to 10$^{22}$
 eV, and consider (with few exceptions) proton propagation for source
 distances less than 200 Mpc. The calculations are carried out using a
 Monte Carlo technique, and we propagate individual protons injected
 as either a mono-energetic beam, or with energies sampled from a
 fixed source energy spectrum. This approach has the
 advantage of representing fluctuations in the proton energy losses
 very well, thereby giving us a good handle on the correlations
 between energy loss, time of flight and angular deviation of the
 flight direction. As we will show, these important UHECR
 characteristics are deeply interconnected.  For a given source
 distance, there is a strong correlation between the amount of energy
 lost, the time delay, and the scattering angle.

 Our calculations are thus mainly relevant to scenarios of UHECR
 acceleration at astrophysical shocks, for which 10$^{22}$ eV is a
 very generous upper energy limit. With this paper we wish to
 establish limits for the distance of potential UHECR proton sources
 as a function of proton energy and the average strength of the
 extragalactic magnetic field. We also study the angular distribution
 of UHECR with respect to the source direction (arrival angle) and the
 time delays after propagation over different distances.  In addition,
 the neutrino fluxes produced during the propagation are presented.

 The article is organized as follows. We describe the propagation
 method, including the relevant features of the event generator
 SOPHIA, in Section 2. Section 3 gives some interesting results on the
 propagation of mono-energetic proton beams, and compares our results
 with other work. Section 4 analyzes the formation and development of
 the primary and secondary particle spectra for protons injected with
 a power law spectrum. In section 5 we discuss the results, present our
 conclusions, and make suggestions for future work.

%%%%%%%%%%%%%%%%%%%%%%%%%%%%%%%%%%%%%%%%%%%%%%%%%%%%%%%%%%%%%%%%%%%%%%%%%%

\section{Cosmic Ray Propagation}

This section provides a description of our simulation code for
propagating protons in intergalactic space. We treat energy losses due
to hadronic and electromagnetic interactions of the nucleons with
photons of the cosmic microwave background radiation as well as the
deflection of particles by the intergalactic magnetic field. Although
we present here only results on nucleon propagation in random magnetic
fields, our approach also allows us to follow the particles in
complicated magnetic field topologies. Because of the time-consuming
detailed simulation of each nucleon propagation path by Monte Carlo,
the propagation method described below is not suitable for calculations
involving large cosmological distances.

%-----------------------------------------------------------------------

\subsection{Interactions and energy loss processes}

Particles of energy $E>10^{18}$~eV interact with photons of the cosmic
microwave background radiation giving rise to secondary particle
production and nucleon energy loss. The most important processes are:
\begin{itemize}
\item photoproduction of hadrons, and
\item Bethe-Heitler (BH) production of $e^+ e^-$ pairs by protons.
\end{itemize}
We also account for the adiabatic losses due to cosmological expansion
of the Universe, and for the decay of neutrons produced in hadronic
production process. Since we restrict our calculation to models of
UHECR acceleration in astrophysical shocks, and energies below
10$^{22}$ eV, we consider only interactions with cosmic microwave
background photons. The calculation of nucleon propagation at higher
energies would require the use of models of the radio background (see
e.g. Ref.~\cite{ProthBier}).  Since we are not presenting results
on the development of electromagnetic cascades initiated by secondary
particles produced in proton-photon interactions, we can safely
neglect interactions on the universal optical/infrared background as
well. We keep track, however, of the individual energies of all
secondaries of photoproduction interactions and are thus able to show
the spectra of neutrinos generated by primary protons after propagation
over different distances.

Hadron production and energy loss in nucleon-photon interactions is
simulated with the event generator SOPHIA~\cite{SOPHIA}.  
This event generator samples collisions of nucleons with photons 
from isotropic thermal or power law energy distributions, using 
standard Monte Carlo techniques.  In this paper the code
has been used with a blackbody spectrum with $T=2.726$ K to represent
the
cosmic microwave background. According to the respective partial cross
sections, which have been parametrized using all available accelerator
data,
it invokes an interaction either via baryon resonance excitation,
one-particle
$t$-channel exchange (direct one-particle production), diffractive
particle
production and (non-diffractive) multiparticle production using string
fragmentation. The distribution and momenta of the final state particles are
calculated from their branching ratios  and
interaction kinematics in the center-of-mass frame, 
and the particle energies and angles in the lab.
frame are calculated by Lorentz transformations. The decay of all unstable
particles except for neutrons is treated subsequently using standard
Monte Carlo methods of particle decay according to the available phase
space.  The neutron decay is
implemented separately into the present propagation code.
The SOPHIA event generator has been tested and shown
to be in good agreement with available accelerator data.
A detailed description of the code including the sampling methods, the
interaction physics used, and the performed
tests can be found in Ref.~\cite{SOPHIA}.

The Monte Carlo treatment of photoproduction is very
important, because nucleons lose a large fraction of their energy
 in each interaction.  As early as 1985 Hill \& Schramm \cite{Hill85}
 pointed out that the use of a continuous energy loss approximation
 for this process neglects the intrinsic spread of arrival energies
 due to the variation of the energy loss $\Delta E$ per interaction,
 and the Poissonian distribution in the number of pion production
 interactions during propagation. This results in a certain ``survival
 probability'' of cosmic rays arriving at Earth with energies {\em above}
 the GZK-cutoff, as estimated in the assumption of continuous
 energy loss.

\begin{figure}[t] % fig 1
\centerline{\epsfig{file=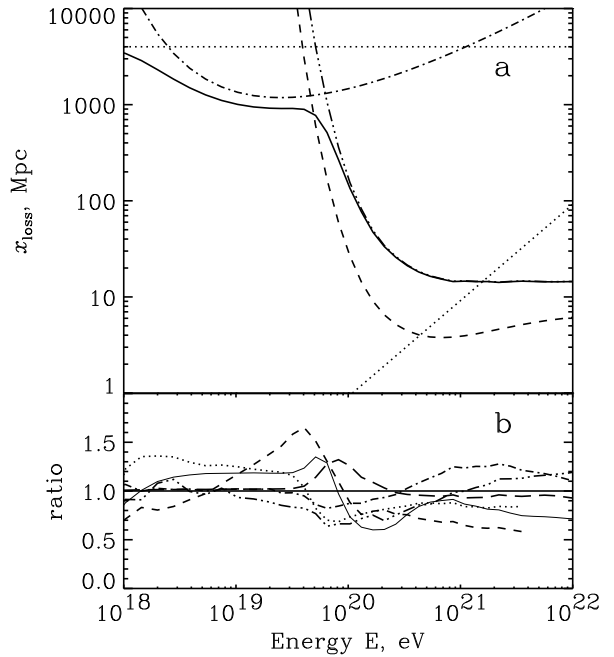,width=80mm}}
\vspace{10pt}
\caption{ a) Mean energy loss length due to adiabatic expansion
(upper dotted curve), Bethe-Heitler pair production (dash-dotted
curve), hadron production (triple-dot-dashed curve). Also shown are the
hadron interaction length (dashed curve) and the neutron decay length
(lower dotted curve). The solid line shows the total \protect$x_{\rm loss}$.
 b) Ratio of mean energy loss length
 as calculated in  Refs.~\protect\cite{BG88} (dotted),
\protect\cite{RB93} (long-dashed),
\protect\cite{YT93} (short-dashed),
\protect\cite{PJ96} (dash-dotted),
\protect\cite{Lee98} (dashed-dot-dot-dot), and
\protect\cite{Achterror} (thin solid) to the loss length
of the present work presented in the upper panel.
\label{fig1}
}
\end{figure}

Fig.~\ref{fig1}a shows the energy dependence of all parameters
relevant to the average proton energy loss in the microwave
background (T=2.726~K) for redshift $z$ = 0. The photoproduction
interaction length $\lambda_{\rm ph}$ for protons is shown as a dashed line.
Denoting the proton-photon center-of-mass energy by $\sqrt{s}$, 
the interaction length can be written as~\cite{PJ96}
\begin{eqnarray}
\frac{1}{\lambda_{\rm ph}(E)} &=&
\nonumber\\
& &\hspace*{-1cm}
\frac{1}{8E^2\beta} \int_{\epsilon_{\rm th}}^{\infty}
d\epsilon \frac{n(\epsilon)}{\epsilon^2} \int_{s_{\rm min}}^{s_{\rm
max}(\epsilon,E)} ds
(s-m_p^2 c^4) \sigma_{p\gamma}(s)
\end{eqnarray}
with 
\begin{eqnarray}
s_{\rm min} &=& (m_p c^2+m_{\pi^0} c^2)^2 
\\
s_{\rm max}(\epsilon,E)&=&m_p^2c^4+2E\epsilon(1+\beta)
\\
\epsilon_{\rm th} &=& \frac{s_{\rm min}-m_p^2 c^4}{2E(1+\beta)}, 
\hspace*{1cm}
\beta^2 = 1 - \frac{m_p^2 c^4}{E^2}\ .
\end{eqnarray}
Here $E$ ($\epsilon$) is the proton (photon) energy and the
proton and neutral pion masses are $m_p$ and $m_{\pi^0}$, respectively.
The CMB photon density is given by $n(\epsilon)$ in units of cm$^{-3}$
eV$^{-1}$ and 
the photoproduction cross section, $\sigma_{p\gamma}(s)$, is taken from
the parametrization implemented in SOPHIA. 

The mean energy loss distance $x_{\rm{loss}}(E)$, shown in
Fig.~\ref{fig1}a as triple-dot-dashed curve,
is calculated as
\begin{equation}
 x_{\rm loss}(E) \; = \; {{E} \over {dE/dx}} \; = \; {{\lambda(E)}
 \over {\kappa(E)}}
%%\frac{1}{E} \frac{dE}{dx_{\rm{loss}}} = \frac{\kappa(E)}{\lambda(E)}
\end{equation}
with $\kappa(E)$ being the mean inelasticity
\begin{equation}
\kappa(E) = \frac{\langle \Delta E \rangle}{E}\ .
\end{equation}
The mean energy loss of the nucleon due to the hadron production, 
$\langle \Delta E \rangle$, has been calculated by
simulating $10^4$ interactions for each given proton energy, resulting in a
statistical error of the order of 1\%.
For $E>10^{20}$~eV losses through photomeson
production dominate with a loss distance of about $15$~Mpc at $E \geq
8\times 10^{20}$~eV. Below this energy, Bethe-Heitler pair production
and adiabatic losses due to the cosmological expansion in the Hubble
flow determine the proton energy losses.

Both the photoproduction interaction and the pair production are
characterized by strongly energy dependent cross sections and
threshold effects.  Fig.~\ref{fig1}a shows $\lambda_{\rm ph}$ decreasing
by more than three orders of magnitude for a proton energy increasing
by a factor of three.  After the minimum $\lambda_{\rm ph}$ is reached,
the proton energy loss distance is approximately constant. It is worth
noting that the threshold region of $\lambda_{\rm ph}$ is very important
for the shape of the propagated proton spectrum. As pointed out by
Berezinsky \& Grigoreva~\cite{BG88}, a pile--up of protons will be formed
 at the intersection of the photoproduction and pair production energy loss
distances. Another, smaller pile--up will develop at the intersection
of the pair production and adiabatic loss functions.

In the current calculation we treat pair production as a continuous
loss process which is justified considering its small inelasticity of
$2m_e/m_p\approx 10^{-3}$ (with $m_e$,$m_p$ being the electron and
proton masses, respectively) compared to pion-photoproduction ($\kappa
\approx 0.2-0.5$). We use the analytical fit functions given by
 Chodorowsky {\em et al.}~\cite{Chodo} to
calculate the mean energy loss distance for Bethe-Heitler pair
production. This result is in excellent agreement with results
obtained by simulating this process via Monte Carlo as done by
Protheroe \& Johnson~\cite{PJ96}.

%%%%%%%%%%%%%%%%%%%%%%%%%%%%%%%%%%%%%%%%%%%%%%%%%%%
%%%%%%%%%%%%%%%%%%%%%%%%%%%%%%%%%%%%%%%%%%%%%%%%%%%
The turning point from pion production loss dominance to pair
production loss dominance lies at $E \approx 6 \times 10^{19}$~eV,
with a mean energy loss distance of $\approx 1$~Gpc. The minimum of
the pair production loss length is reached at $E \approx (2-4) \times
10^{19}$~eV.  For $E\leq (2-3) \times 10^{18}$~eV continuous losses
due to the expansion of the universe dominate. For an Einstein-de
Sitter (flat, matter-dominated) universe as considered here, the
cosmological energy loss distance scales with redshift $z$ as
\begin{equation}
 x_{\rm{loss,ad}}(E,z) = \frac{c}{H_0} (1+z)^{-3/2}
 \approx 4000~{\rm Mpc}~ (1+z)^{-3/2} ,
\end{equation}
for a Hubble constant of $H_0=75$~km/s/Mpc, which we use throughout
this paper.  All other energy loss
distances, $x_{\rm{loss,BH}}$ for Bethe-Heitler pair production and
$x_{\rm{loss, ph}}$ for photomeson production, scale as
\begin{equation} x_{\rm{loss}}(E,z) = (1+z)^{-3}
x_{\rm{loss}}[(1+z)E,z=0]\ .
\end{equation}

We also show the mean decay distance of $\sim 9\times 10^{-9}
\gamma_n$~kpc for neutrons, where $\gamma_n$ is the Lorentz factor of the
neutron. Obviously, neutrons of energy below 10$^{21}$ eV
tend to decay, whereas at higher energies neutrons tend to interact.

 Since the details of the proton energy loss directly affect the
 proton spectra after propagation, we present the ratio of the loss
 distance in previous calculations to that of our work on a linear
 scale in Fig.~\ref{fig1}b. Generally all values of the energy loss
 distance are in a good qualitative agreement. 
 Rachen \& Biermann~\cite{RB93} treat both Bethe-Heitler
 and pion production losses very similarly to our work except for the
 threshold region of pion production. 
 In the pair production region our work is also in perfect agreement
 with Protheroe \& Johnson~\cite{PJ96}. An overestimate of the loss
 distance due to pion production of $\sim 10-20\%$ in Ref.~\cite{PJ96},
 however, will result in a small shift of the GZK cutoff to
 higher energies in comparison to the present calculations. 
 Berezinsky \& Grigoreva~\cite{BG88} used a very good approximation
 for the pion production losses, but underestimate the energy loss in
 pair production interactions by at least 30-40\%. The largest deviation
 of the combined loss distance from our model appears in the calculations
 of Yoshida \& Teshima~\cite{YT93}. As already pointed out in
 Ref.~\cite{PJ96} the largest difference occurs at
 $\approx 5 \times 10^{19}$ eV
 where Ref.~\cite{YT93} underestimates pair production losses and
 uses $x_{\rm loss}$ values larger by about 60\%, while  photoproduction
 losses are overestimated by up to 50\%.  In the  work of
 Lee~\cite{Lee98} pion as well as pair production losses are treated in
 fair agreement with our work, with differences up to 40\% 
 in the threshold region of pion production, and 10-20\% otherwise.
The energy loss code of Lee was also used by Sigl and collaborators
\cite{Sigl97a,Sigl99a}.
 The simple analytical estimate of photoproduction losses in
 the recent work of Achterberg {\em et al.}~\cite{Acht99,Achterror}
 underestimates the photoproduction loss distance by $10 {-} 40 \%$,
 while $x_{\rm loss}$ due to pair production losses is overestimated
 by about $20\%$.

%------------------------------------------------------------------------

\subsection{Method of particle propagation}

UHECR propagation involves two main distance scales: (a) the hadronic
interaction length $\lambda_{\rm ph}$ of typically 3 to 7 Mpc, and (b)
the much smaller length scale $\ell_{\rm mag}$ of typically 10 kpc
needed for a precise numerical integration of the equations of motion
in a random magnetic field. A straightforward Monte Carlo treatment of
the propagation using a step size of $\ell_{\rm mag}$ for both
hadronic interactions and the equations of motion leads to severe
efficiency problems for total propagation distances of hundreds of
Mpc.  Hence, the Monte Carlo simulation is done in the following
way. First the path length $X_{\rm dist}$ from the current particle
position to the next possible hadronic interaction is determined from
\begin{equation}
X_{\rm dist} = - \lambda_{\rm ph,min} \ln(\xi)\ ,
\label{int-length}
\end{equation}
where $\lambda_{\rm ph,min}$ is the minimum interaction length for
hadronic interactions (at maximum redshift possible for a given total
propagation distance) and $\xi$ is a random number uniformly
 distributed in $(0,1]$. The nucleon is then propagated over the path
 length $X_{\rm dist}$ in steps of $\ell_{\rm mag}$, and for charged
 particles Bethe--Heitler losses are taken into account and
 the deflection angle is calculated.
 A hadronic interaction is then simulated with the
 probability $\lambda_{\rm ph,min}/\lambda_{\rm ph}(E,z)$,
 $\lambda_{\rm ph}(E,z)$ being the interaction length for the
 energy $E$ and redshift $z$.  It is shown in Appendix A that this method
 corresponds exactly to a propagation simulation using
 Eq.~(\ref{int-length}) with $\lambda_{\rm ph}(E,z)$ for the
 calculation of the interaction distance at each step with the length
 $\ell_{\rm mag}$.

 The reduction of the proton energy due to BH pair production and
 of all nucleons due to adiabatic expansion is calculated at every
 propagation step, whereas the corresponding loss lengths are updated
 after a simulated path length of $\lambda_{\rm ph,min}$ and every
 photoproduction interaction.  In the case of neutrons the decay
path length is sampled using Eq.~(\ref{int-length}) with the neutron
decay length. The smaller of both the hadronic interaction and the
decay lengths determines then the larger scale of the simulation.

If a photoproduction interaction has occurred, the new energy of the
proton (neutron) is substituted for the old one, and the energies and
particle types of the secondary particles are recorded.  The event
generator SOPHIA generates the full set of secondary particles,
including nucleon--antinucleon pairs. Thus the total flux of nucleons
after propagation is slightly higher than the injected proton
flux. Although this is not essential for the main results of this
paper, it may occasionally affect the normalization of the proton
arrival spectra.

 The propagation is completed when the distance between the injection
 point and the particle location exceeds the predefined source
 distance. To obtain precise results for the time delay (e.g. total
 nucleon path length compared to the path length of a light ray), the
 last integration step is adjusted to end exactly at the desired
 distance.
% This may lead to an underestimate of the time delay when the particle
% fluxes become nearly isotropic and many particles have a high probability
% to scatter back through the `observer's sphere'. It will not, however,
% affect strongly the results presented in this paper.

 Particles are injected at a point in space with a randomly chosen
 small angular deviation from the $z$-axis which defines the main
 propagation direction. The space along the $z$-axis is subdivided into
 $32\times32\times512$ cubes of side 250 kpc, each filled with a
 random magnetic field
 of average strength $\langle B \rangle$ = 10$^{-9}$ Gauss (1
 $n$G)~\cite{Kronberg94} satisfying a Kolmogorov spectrum with three
 logarithmic scales.
 In practice three field vectors of random orientation are sampled at
 scales $\ell$ = 1000, 500, and 250~kpc with amplitudes proportional
 to $\ell^{\; 1/3}$ (see Appendix B).
 The final magnetic field in each of the 250 kpc cubes is the vectorial
 sum of these three vectors.
 Cyclic boundary conditions are imposed in case a
 particle leaves the space of pre-calculated magnetic fields.  
This means that the magnetic field experienced by a particle at 
location  $\bf x$ is the same as the field calculated at
$\bf x^\prime$,
\begin{equation}
x_i^\prime = x_i - N_i R_i,\hspace*{2cm}i=x,y,z
\end{equation}
with $R_i$ being the size of the pre-calculated magnetic field region
in direction $i$. $N_i$ is the largest integer number satisfying
$x_i-N_i R_i \ge 0$.
The
 magnetic field values are refreshed after the calculation of 100
 propagations to exclude systematic effects by our choice
 of field vectors.
 We have verified numerically that the magnetic field constructed in
 this way obeys approximately div(${\mathbf B}$) = 0 and
 that recalculations of the field at smaller intervals do not change
 the final result. We assume that the magnetic field strength does
 not scale with redshift. More information about the implementation
 of the random magnetic field is given in Appendix B.

The value chosen for $\ell_{\rm mag}$, in principle, depends strongly
on the average magnetic field and nucleon energy, and is a
compromise between the precision of the calculation and computing time
limits. We have chosen $\ell_{\rm mag}$ = 10 kpc for $\langle B
\rangle$ = 1 $n$G, with an inverse linear scaling for other $B$
values. A step size of 1 kpc has been used for short distance
propagations to ensure accurate results for arrival angle and time
delay distributions.

Finally, it should be mentioned that the calculation of the redshift
at a given distance can be done only approximately. The reason is the
unknown total travel time of a particle from the source to Earth at
injection time.  The actual travel time (path length) can be
significantly larger than the light travel time along a geodesic and
is, in general, different for each simulated particle trajectory. In
the following we use the proper distance-redshift relation to define
the redshift of the source and along the travel path at observation
time. This approximation does not strongly affect our results since we
consider here mainly distances with redshifts smaller than 0.06 and weak
magnetic fields. However, it should be noted that, in the case of a strong 
magnetic field, cosmological evolution might become important already 
at relatively short distances.

%%%%%%%%%%%%%%%%%%%%%%%%%%%%%%%%%%%%%%%%%%%%%%%%%%%%%%%%%%%%%%%%%%%%%%%%%%

\section{Results and comparison with previous work}

In this section, we present results from the simulation of proton
propagation.  We start with mono-energetic proton fluxes for which we
can compare our results with previous work, and which reflect more
directly the different treatments of the energy loss processes. We
then compare results for the propagation of protons injected with a
power law spectrum.

One can divide previous calculations into two general groups:
Monte Carlo based methods, like our own one, and analytical/numerical
calculations.  Protheroe \& Johnson~\cite{PJ96} have used a matrix
technique to follow the particles over cosmological distances and
calculate the $\gamma$-ray, neutrino and nucleon spectra arriving at
Earth. The energy loss matrices for all particles are calculated with
Monte Carlo event generators.  We have compared our SOPHIA event
generator with the one of Ref.~\cite{PJ96} by propagating
 with the same method an $E^{-2}$ proton spectrum with different
 exponential cutoffs (see Eq.~(\ref{spectrum})). For this purpose we
 have used SOPHIA and the event generator of Ref.~\cite{PJ96}
to calculate the corresponding photoproduction matrices and have
applied the two matrices to propagation over the same set of
distances. A comparison of the resulting secondary
particle spectra yields excellent agreement, pointing to a similar
treatment of the particle production process in the different codes.
We have also compared the matrix method with our Monte Carlo approach
by propagating an exponentially modified power law injection spectrum
over 200~Mpc. Again good agreement is found for the resulting
$\nu_{\mu}$-spectra, while the $\bar\nu_e$- and neutron spectra are at
variance with our calculations, which we attribute to a different
treatment of the neutron decay.
 Also, our Monte Carlo method results in more losses due to pair
production for distances $\geq 200$~Mpc and a sharp spike at the
injection energy for very short distance propagation, a consequence of
the Poisson nature of photon-proton encounters. This feature is
 discussed in detail in Sect.~III.A.

 The approach used by Berezinsky \& Grigoreva~\cite{BG88} and Rachen \&
 Biermann~\cite{RB93} is to solve the transport equation
 quasi-analytically by approximating the collisional terms
 as continuous energy loss terms.
 This does not take into account the Poissonian nature of the
 pion production process as pointed out above, and introduces
 artifacts into the resulting nucleon spectra in form of sharp
 pile-ups. Lee~\cite{Lee98} used a numerical technique to solve the
 transport equation for particle propagation without using the
 continuous loss approximation.

 The common assumption in all this work is to consider the spatial
 propagation as strictly along a null-geodesic, with the consequence
 of not being able to gain knowledge about time delays and arrival
 angles of the cosmic rays with respect to light and neutrino
 propagation.

 A hybrid model, combining a Monte Carlo particle transport code
 with analytical techniques was presented by Achterberg
 {\em et al.}~\cite{Acht99}. Besides simplifying the properties 
 of the energy losses by analytical estimates (see Fig.~\ref{fig1}b),
 this code also describes the scattering in the magnetic field as
 a diffusion process employing stochastic differential equations.
 This approach has the advantage to allow large propagation steps,
 and is thus computationally very fast, but has a disadvantage
 at small propagation distances which we discuss further below.
 Our approach is to use the Monte Carlo
 technique for simulating particle production and to follow closely
 cosmic ray orbits in 3D-magnetic field configurations while traveling
 through the nearby Universe to Earth. This concept,
 while being the most accurate one, limits our propagation
 calculation to small source distances.

%------------------------------------------------------------------------

\subsection{Propagation of mono-energetic protons}

 In this section we present distributions of arrival energy, arrival
 direction and time delay of the nucleons, as well as neutrino
 spectra, for mono-energetic injection of protons at distances of 2,
 8, 32, 128 and 512~Mpc from Earth. Protons are injected with energy
 $10^{21.5}$~eV. At this energy, propagated protons can easily suffer
 several photoproduction interactions, and this tends to emphasize the
 pion production features.

%%%%%%%%%%%%%%%%%%% ARRIVAL ENERGY DISTRIBUTION %%%%%%%%%%%%%%%%%%%%%%%%%%%%

\begin{figure}[tb] % fig 2
\centerline{\epsfig{file=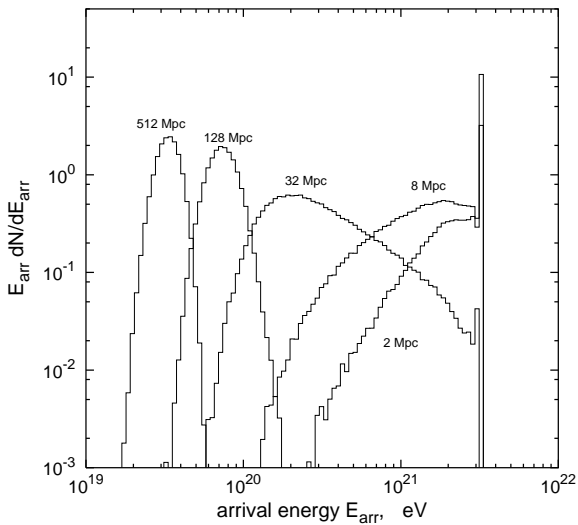,width=85mm}}
\vspace{10pt}
\caption{Arrival spectrum at Earth for mono-energetic injection of protons
of energy $E=10^{21.5}$~eV and for various source distances as indicated.
The sharp spike at injection energy for distances $D\leq 32$~Mpc is due to
the low interaction probability within the short distance.
\label{fig2}
}
\end{figure}

 Fig.~\ref{fig2} shows the distribution of arrival energy of protons
 and neutrons. Clearly visible is the effect of the statistical nature
 of photon--proton encounters, also found qualitatively in Ref.~\cite{Acht99}.
  At a distance of 2~Mpc, roughly 60\% of all injected
 particles do not interact, and this generates a sharp spike at the
 injection energy. This effect due to Poisson statistics remains
 visible for distances up to $\sim 30$~Mpc, showing up as a
 high--energy spike in the cosmic ray
 spectrum.  At larger distances, essentially all injected particles
 undergo interactions, and therefore, the high-energy spike vanishes.
 The arrival energy distributions then become much narrower, and in
 propagation over larger distances would scale simply with the energy
 loss distance for pair production and adiabatic losses, modified by
 the increasing scattering in the magnetic field.
%%%%%%%%%%%%%%%%%%%%%%%%%%%%%%%%%%%%%%%%%%%%%%%%%%%%%%%%%%%%%%%
%%%%%%%%%%%%%%%%%%%%%%%%%%%%%%%%%%%%%%%%%%%%%%%%%%%%%%%%%%%%%%%

%%%%%%%%%%%%%%%% TIME DELAY DISTRIBUTION %%%%%%%%%%%%%%%%%%%%%%%%%%%%%%%%%%%%%%%

\begin{figure}[tb] % fig 3
\centerline{\epsfig{file=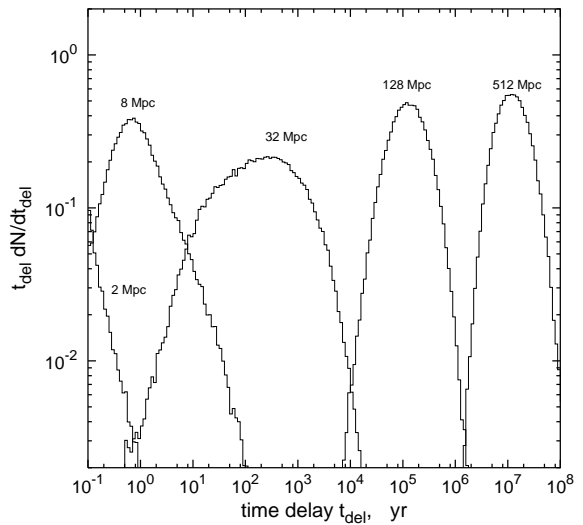,width=85mm}}
\vspace{10pt}
\caption{Time delay of protons injected at different source distances
and propagated through a random magnetic field of 1 $n$G. The time delay
is defined as the propagation time of a particle minus the travel time
of a light ray along a geodesic.
\label{fig3}
}
\end{figure}

 Fig.~\ref{fig3} shows the distribution of the average time delay of
 the cosmic rays arriving at Earth with respect to propagation along a
 geodesic with the speed of light. This delay is caused by scattering
 of the charged particles by the intergalactic magnetic field, leading
 to an increase of the particle's effective path length. Thus, the
 average time delay increases with propagation distance, as visible in
 Fig.~\ref{fig3}.  Like the arrival energy distributions, the
 distributions of the time delay also show signs of Poisson
 statistics, visible especially when propagating over short
 distances.
 
% The time delay effectively reflects the arrival energy
% distribution $t_{\rm del}  \sim 1/E_{\rm arr}^2$ 
% as a result of the random walk process~\cite{Clay98a,Acht99}.
The time delay effectively reflects the arrival energy distribution
$
t_{\rm del} \propto 1/E_{\rm arr}^2$ as a result of the random walk
process~\cite{Clay98a,Acht99}. 
This also emphasizes the importance of an accurate treatment of
energy losses. For example, a direct comparison with the propagation
code of Achterberg {\it et al.}~\cite{Acht99,Achterror} 
for (almost) the same propagation parameters has shown
differences in the time delay up to one order of magnitude for $D=32$
Mpc. For the same propagation distance, the code by Achterberg {\it et
al.} produces a peak in the arrival energy distribution about a factor of
2 lower than found in the present work, due to its $20\%$ overestimation
of energy losses in the photoproduction regime. Together with a difference
in the magnetic field sampling, which leads to an effective correlation
length
$\ell_{\rm corr}\approx 390$ kpc for the Kolmogorov spectrum used in the
present work (see Appendix B) compared to $\ell_{\rm corr} = 1$ Mpc for
the
homogeneous cell approach used in Ref.~\cite{Acht99}, the observed differences
can then be fully understood by the relation $t_{\rm del} \propto \ell_{\rm
corr}/E_{\rm arr}^2$, as derived in Ref.~\cite{Acht99}.

 Protons with injection energy $\leq 10^{19}$~eV suffer mainly
 continuous BH pair production and adiabatic losses that are
 proportional to their path length. The substantial deflection
 in the random magnetic field at such energies results in
 a significant increase of the path length. For protons injected
 at a sufficiently large distance this can also lead to excessive
 time delays. For example, cosmic rays with energy of about
 $10^{19}$~eV, injected at distances greater than $500$~Mpc in a
 1~$n$G magnetic field, show a time delay exceeding the Hubble time.
 This gives a strict constraint on the cosmic ray horizon.
%%%%%%%%%%%%%%%%%%%%%%%%%%%%%%%%%%%%%%%%%%%%%%%
%%%%%%%%%%%%%%%%%%%%%%%%%%%%%%%%%%%%%%%%%%%%%%%

%%%%%%%%%%%%%%%%%%%%%%% CORRELATION BETWEEN ENERGY AND DELAY %%%%%%%%%%%%%%%

\begin{figure}[tb] % fig 4
\centerline{\epsfig{file=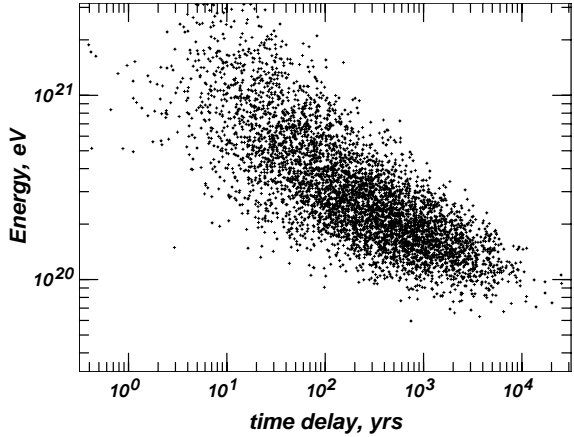,width=85mm}}
\vspace{10pt}
\caption{Scatter plot of time delay versus energy for protons injected
 with energy of 10$^{21.5}$ eV after propagation over 32~Mpc in random
 $B$ field of 1 $n$G.
\label{corr}
}
\end{figure}

 The diffusion coefficient for an effective description of the
 scattering process in the magnetic field is strongly energy
 dependent, and so is the time delay, $t_{\rm{del}}$.
 To emphasize this correlation, and demonstrate the advantages of the
 Monte Carlo approach, we show in Fig.~\ref{corr} the scatter plot of
 proton energy versus delay after propagation over 32~Mpc. There is a
 strong correlation suggesting that energy changes of one and a half
 orders of magnitude lead to differences in delay times of more than
 three orders of magnitude, i.e. we find an energy dependence similar
 to $\langle t_{\rm{del}} \rangle \propto (B\,D/E)^2$ as derived
 by Achterberg {\em et al.}~\cite{Acht99} in the small scattering-angle
 approximation and the quasi-linear approximation of wave-particle
 interactions.  The correlation becomes less pronounced when propagating
 over significantly larger distances simply because the arrival energy
 distributions become much narrower and the statistical nature of the
 energy loss is smoothed by the prevailing pair production and
 adiabatic losses. This correlation, however, would have very
 important implications for specific models of UHECR production, where
 the duration of an active phase of the source competes with the time
 delay of the protons during propagation. The extreme case would be
 the acceleration of UHECR in gamma ray bursts.  The particles with
 the highest energies are expected to arrive first, followed by a
 dissipating widening halo of lower energy protons, as emphasized
 by Waxman \& Miralda--Escud\'e~\cite{WME96}.

% At small propagation distances ($D \le 32$ Mpc), time delays are
% only a very small fraction of the total propagation time. This makes
% their value very sensitive to a precise treatment of the particle motion,
% which becomes obvious when we compare our results to those obtained
% by Achterberg {\em et al.}~\cite{Acht99,Achterror} - their average time
% relays for distances smaller than $100$ Mpc are a factor $10{-}100$
% larger than our values. As noted in Ref.~\cite{Acht99}, their description
% of particle scattering as a diffusion process is {\em not} 
% equivalent to a treatment of detailed particle orbits in a given
% field realization for distances smaller or of order of the magnetic field
% decorrelation length, i.e. $\sim 30$ Mpc for the magnetic field
% implementation in this work and in Ref.~\cite{Acht99,Achterror}.   
% The differences in $t_{\rm del}$ between the two calculations
% can be explained only in part by the stronger energy evolution found
% in the results of Ref.~\cite{Achterror}, and the energy--delay time
% correlations shown above. Most of the effect is probably due to
% problems of the diffusion approach at small propagation
% distances, which emphasizes the importance of detailed
% Monte Carlo simulations in this regime.

 For large propagation distances, even protons injected with
 10$^{21.5}$ eV show time delays that are a considerable fraction
 of the light propagation time ($5-10\%$ for  $512$~Mpc).
 This would lead to a limiting proton horizon for a large set of
 source distances and magnetic field values~\cite{Acht99}.
 512~Mpc is already a limiting horizon for protons injected with
 10$^{19}$~eV in 1~$n$G fields, as noted above.
%%%%%%%%%%%%%%%%%%%%%%%%%%%%%%%%%%%%%%%%%%%%%%%%
%%%%%%%%%%%%%%%%%%%%%%%%%%%%%%%%%%%%%%%%%%%%%%%%

%%%%%%%%%%%%%%%%%%%%%%% ARRIVAL ANGLE DISTRIBUTION %%%%%%%%%%%%%%%%%%%%%%%%%%%

%%%%%%%%%%%%%%%%%%%%%%%%%%%%%%%%%%%%%%%%%%%%%%%%
\begin{figure}[tb] % fig 5
\centerline{\epsfig{file=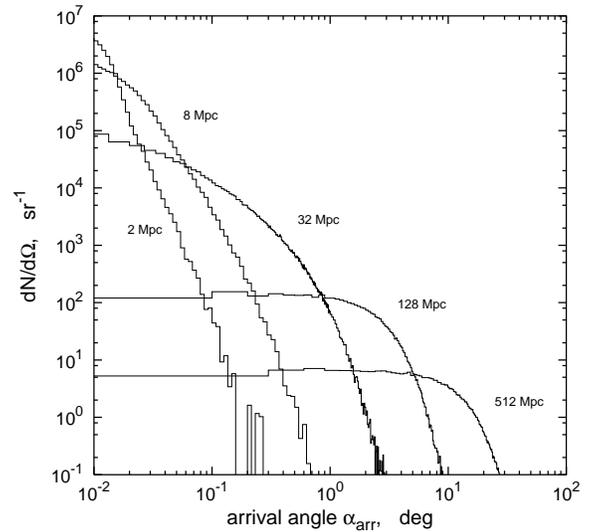,width=85mm}}
\vspace{10pt}
\caption{Angular distribution of the arrival angle at Earth for
 mono-energetic injection of protons
of energy $E=10^{21.5}$~eV, and for various source distances as indicated.
The magnetic field is 1~nG.
\label{fig5}
}
\end{figure}
%%%%%%%%%%%%%%%%%%%%%%%%%%%%%%%%%%%%%%%%%%%%%%%%
 The scattering that leads to time delay also causes angular
 deviations from the direction to the source, as shown in
 Fig.~\ref{fig5} for the injection of mono-energetic protons at the same
 set of distances.  Note that in our propagation code the `observer' sits
 on a sphere surrounding the injection point. The angle shown is the
 angle between the particle's arrival direction and direction to the
 injection point.  This `arrival angle' is somewhat different from the
 angle between particle's arrival direction and the injection
 direction. This method may lead to an underestimate of the scattering
 angle and the time delay when the particle fluxes become nearly
 isotropic and many particles have a high probability to scatter back
 through the `observer's sphere'. It will not, however, affect strongly
 the results presented in this paper, because, as Fig.~\ref{fig5}
 demonstrates, we do not reach the limit of isotropic 3D diffusion.
 
 The features of the angular distribution closely follow the
 time delay distributions already shown.
 For large propagation distances, the cosmic ray arrival directions
 are distributed uniformly up to a maximum deflection angle, which
 increases with propagation distance to reach more than 20$^\circ$ at
 512~Mpc. 
 At propagation distances smaller than $\sim$30~Mpc, thus a few times the
 proton interaction length $\lambda_{\rm ph}$, a peak at small deflection
 angles occurs due to the effect of Poisson statistics for proton--photon
 interactions.
%%   Propagation over small distances
%%generate wider distributions because of the effects of Poisson statistics
%%for photon--proton interactions. At distances somewhat
%%larger than the proton interaction length $\lambda_{\rm ph}$, when almost
%%all protons suffer photoproduction interactions, the angular distribution
%%is very wide, reflecting the energy spread of the secondary nucleons.
%%The maximum deflection angle $\alpha_{\rm arr}$ increases with propagation
%%distance to reach more than 20$^\circ$ at 512 kpc.

%%At large distances, where continuous energy losses prevail,
%%the angular distributions are again quite narrow.

%%%%%%%%%%%%%%%%%%%%% NEUTRINO SPECTRA %%%%%%%%%%%%%%%%%%%%%%%%%%%%%%%%%%%%%

%%%%%%%%%%%%%%%%%%%%%%%%%%%%%%%%%%%%%%%%%%%%%%%%%%%%
\begin{figure}[tb] % fig 6
\centerline{\epsfig{file=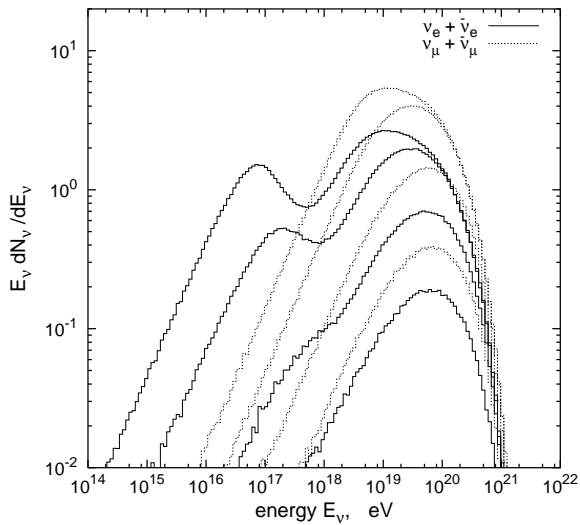,width=85mm}}
\vspace{10pt}
\caption{$\nu_\mu+\bar\nu_{\mu}$ and $\nu_e+\bar\nu_e$-spectra at
 Earth after propagating a mono-energetic proton beam of energy
 $10^{21.5}~$eV at distances of 2, 8, 32 and 128~Mpc (from bottom
 to top) in a 1~\protect$n$G intergalactic magnetic field.
\label{fig6}
}
\end{figure}
%%%%%%%%%%%%%%%%%%%%%%%%%%%%%%%%%%%%%%%%%%%%%%%%%%%%%
 Finally Fig.~\ref{fig6} shows the electron and muon neutrino spectra
 generated by the injection of 10$^{21.5}$ eV protons at the same set
 of distances. The muon neutrino spectra develop as a function of
 the proton arrival energy spectra folded with the photoproduction
 cross section. The fluxes grow with propagation distance, and the
 maximum neutrino energy shifts to lower energy reflecting the
 decreasing proton energy. The growth rate with
 distance decreases for very large distances, where the average
 proton energy significantly decreases and $\lambda_{\rm ph}$
 is correspondingly significantly longer.

 Electron neutrino spectra show another, very interesting feature,
 that develops with distance. At a minimum distance of 2~Mpc the
 $\nu_e$-flux reaches its maximum of 1/2 of the $\nu_\mu$ spectrum and
 shows a somewhat wider energy spectrum, enhanced at low energy. At
 larger distances an additional $\nu_e$ component develops at
 significantly lower energy.  As already noted in Ref.~\cite{YT93},
these are $\bar{\nu}_e$'s from neutron
 decay.  The resulting protons from the decay process carry most of
 the energy, leaving for the $\bar{\nu}_e$'s an average energy of only
 $\approx 5 \times 10^{-4}$ of the original neutron energy, and the
 $\nu_e$-peak is placed at about two orders of magnitude to lower
 energy with respect to the $\nu_\mu$-peak.  The strength of this
 component increases with distance relative to the direct $\nu_e$
 component from $\mu^\pm$ decay.

%-------------------------------------------------------------------------

\subsection{Cosmological modification of the cosmic ray source spectrum}

 Berezinsky \& Grigoreva~\cite{BG88} introduced the modification
 factor $M(E,z)$ to represent the cosmological evolution of the UHECR
 spectra. $M(E,z)$ gives the ratio of propagated to injected protons
 at the same energy $E$, for a fixed injection spectrum, as a function of
%the arrival proton energy $E_{\rm arr}$ and
 the redshift of the injection distance compensating for the proton
 adiabatic losses. $M(E,z)$ is thus exactly unity for proton energies
 below the $p\gamma$-particle production energy threshold.

\begin{figure}[tb] % fig 7
\centerline{\epsfig{file=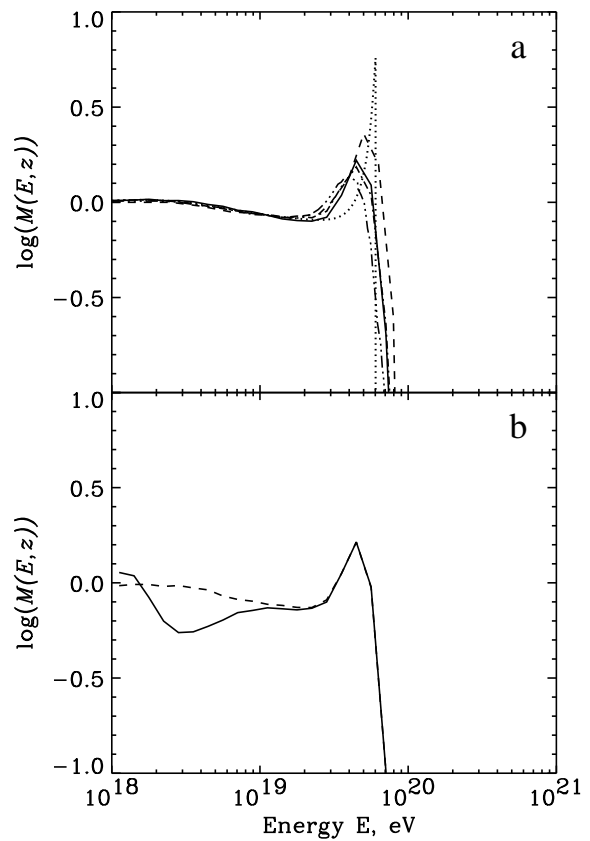,width=75mm}}
\vspace{10pt}
\caption{Upper panel: modification factors for propagation over a
distance of 256~Mpc without magnetic field after injection of a
\protect$E^{-2}$ proton spectrum with a sharp
 cutoff at \protect$E_c=3\times 10^{20}$~eV. This calculation
 (solid line) is compared
to Refs.~\protect\cite{RB93} (dotted line, D= 240~Mpc),
\protect\cite{PJ96} (dashed-dotted line, D=256~Mpc),
\protect\cite{YT93} (dashed line, D=228~Mpc, \protect$E_c=10^{20}$~eV) and
\protect\cite{Lee98} (dashed-dot-dot-dot line, D=256~Mpc).
 Lower panel: comparison of the modification factor for
rectilinear propagation (dashed curve, 'no scattering' curve) and for
propagation in a 1~\protect$n$G magnetic field
(solid line, 'scattering curve') including
the effect of scattering.
\label{fig7}
}
\end{figure}

 At the highest injection energies the modification factor shows the
 GZK cutoff, followed by a pile--up at the crossover of
 photoproduction and pair production energy loss. This pile--up is a
 direct consequence of the resonance nature of photoproduction and the
 hadronic particle production threshold. The next feature at still
 lower energy is a shallow dip corresponding to the pair production
 loss, followed by a small pile--up below it. The magnitude of the
 pile--ups and dips depend not only on the distance and the mean loss
 distance at the photoproduction/pair production crossover, but also
 on the shape of the proton injection spectrum. Flatter spectra create
 bigger pile--ups, because of the increased number of higher energy
 protons that have interacted to lose energy.  The pile--up energy is
 linked to the energy where losses due to pair production take over
 from pion production losses, and is therefore strongly dependent on
 the details of the loss processes in the simulations.
 Fig.~\ref{fig7}a shows $M(E,z)$ for propagation without magnetic
 field for the sole reason of comparison with previous work.  An
 $E^{-2}$ proton spectrum with a sharp cutoff at $E_c=3\times
 10^{20}$~eV is injected, and we propagate over a distance of 256~Mpc
 in our calculation (solid line) compared to Refs.~\cite{RB93} (dotted
 line, $D$=240~Mpc), \cite{PJ96} (dashed-dotted line, D=256~Mpc),
 \cite{YT93}(dashed line, $D$=228~Mpc, $E_c=10^{20}$~eV) and
 \cite{Lee98} (dashed-dot-dot-dot line, $D$=256~Mpc).  
There is excellent agreement at all energies with the work of Protheroe
\& Johnson \cite{PJ96}.  The sharp
 photoproduction peak of Rachen \& Biermann~\cite{RB93} is an artifact
 coming from their continuous loss approximation for pion
 photoproduction. As noted previously, Yoshida \& Teshima~\cite{YT93}
 used a loss curve which shows a significant deviation from that used
 in the present paper, and hence their corresponding pile--up height
 is also larger than in our work.  
 We agree with the position of the pile--up of Lee~\cite{Lee98}.
%% In the recent work of Lee~\cite{Lee98}
%% this pile--up energy is also lower in comparison to Ref.~\cite{PJ96}.
 However, due to an overestimate of the loss rate at this energy,
 the magnitude of the  pile--up in this paper is smaller than in our
 model.  The
 dip just below the pile-up is in reasonable agreement with all other
 works.

%%%%%%%%%%%%%%%%%%%%%%%%%%%%%%%%%%%%%%%%%%%%%%%%%
%%%%%%%%%%%%%%%%%%%%%%%%%%%%%%%%%%%%%%%%%%%%%%%%%
Fig.~\ref{fig7}b illustrates the effect of scattering in the magnetic
field by comparing the resulting corresponding modification factors.
The `no scattering' curve (dashed line, as in Fig.~\ref{fig7}a) is much
higher than the more realistic `scattering curve' in the energy range
between 10$^{18}$ and 10$^{19}$~eV. The reason is that particles in
this energy range have considerable time delays and correspondingly
much higher total energy loss in pair production interactions. Another
consequence of the increased proton travel time due to scattering is
the development of a higher pile--up at about 10$^{18}$~eV,
corresponding to the large number of particles moved to lower energies
from the region of that dip. Note that simulation of 10$^{18}$~eV
particles in a 1~$n$G field is at the threshold of our direct
Monte Carlo approach, and the calculation is not carried to lower
energy where it might show an additional pile--up content.
Fig.~\ref{fig7}b thus demonstrates the importance of the proton
scattering in the extragalactic magnetic fields for the shape of the
final spectrum on arrival at Earth.

It is important to note that the curves shown in Fig.~\ref{fig7}b are
calculated for a source with unlimited lifetime. In addition, by
construction, energy loss due to cosmological evolution does not enter
the modification factor $M(E,z)$. Imposing a constraint on the source
lifetime will change the modification factor considerably for low
energies because, for a given distance, the time delay due to the 
scattering in the turbulent
magnetic field might become comparable to or even exceed the source
lifetime.

%%%%%%%%%%%%%%%%%%%%%%%%%%%%%%%%%%%%%%%%%%%%%%%%%%%%%%%%%%%%%%%%%%%%%%%%

\section{Formation of the primary and secondary particle spectra
 during propagation}

 To study the development of the primary and secondary particle
 spectra we followed the propagation of protons injected with a
 $E^{-2}$ power law spectrum with an exponential cutoff at
 10$^{21.5}$~eV, i.e.
\begin{equation}
 {{dN} \over {dE}} \; = \; A E^{-2} \exp[-E/(10^{21.5} {\rm eV})]\ .
\label{spectrum}
\end{equation}
 We recorded the spectra after propagation over 10~Mpc
 intervals up to a source distance of 200~Mpc.  The results of this
 calculation are relevant for models of UHECR acceleration at
 astrophysical shock fronts, although the cutoff energy
 adopted in this calculation is fairly high. 10,000 protons were
 injected with a power law spectrum (integral spectral index $\gamma$ =
 1) in each of 30 energy bins covering energies from 10$^{19}$ to
 10$^{22}$ eV, i.e. 10 bins per decade of energy.  We did not simulate
 the propagation of lower energy particles, which do not experience
 photoproduction interactions, but followed the secondaries down to
 arbitrary low energies.

\begin{figure}[tb] % fig 8
\centerline{\epsfig{file=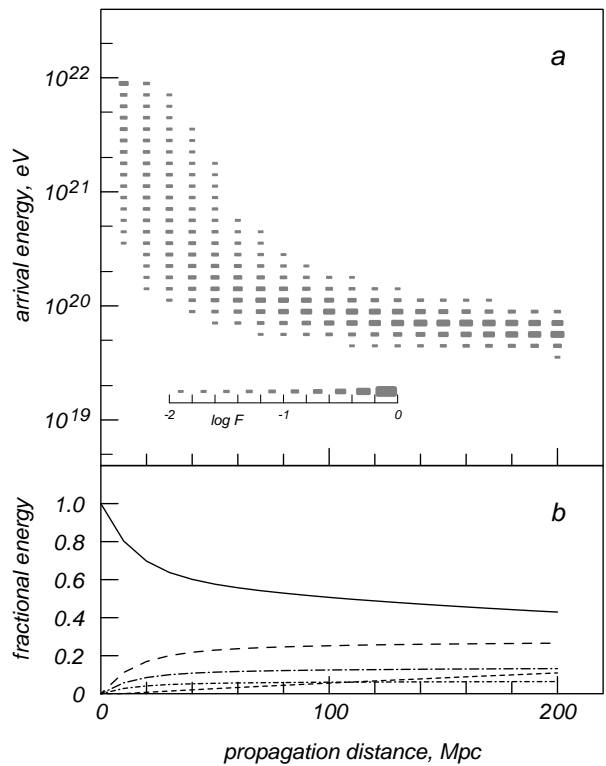,width=85mm}}
\vspace*{10pt}
\caption{ a) Arrival energy distribution for protons injected with
 energy between 10$^{21.9}$ and 10$^{22}$ eV after propagation on
 10, 20, ... 200 Mpc.
 b) Fractional energy contained in nucleons (solid line),
 $\gamma$--rays from photoproduction (long dashes) and BH
 pair production (short dashes) for protons injected with
 the energy spectrum of Eq.~\ref{spectrum}. The dash--dot
 lines show the fractional energy in muon (long) and electron
 (short) neutrinos and antineutrinos. 
\label{power}
}
\end{figure}

  Fig.~\ref{power}a shows the evolution of the particles injected in
 the highest energy bin 10$^{21.9}$ to 10$^{22}$~eV. The size of each
 rectangle is proportional to the fractional energy distribution after
 propagation over 10, 20, etc., Mpc. The rate of energy degradation
 is dramatic. After only 10~Mpc the spectrum of protons injected in a
 0.1 logarithmic bin have spread over one and a half orders of
 magnitude.  The width of the energy distribution increases with the
 propagation distance up to $\sim$30 Mpc and then
 decreases. Qualitatively this behavior is very similar to the
 calculation of Aharonian \& Cronin~\cite{AhCronin}, although the
 direct comparison is difficult because of the different approach to
 the calculation.  The average behavior of all protons injected with
 energy above about $3\times 10^{20}$ eV is similar, although the
 magnitude of the spread decreases --- particles of energy below
 10$^{20}$ eV suffer much smaller losses. After propagation over about
 100~Mpc the spectrum shown in Fig.~\ref{power} is already final - it
 is concentrated within roughly 1/2 order of magnitude around $\sim 8
 \times 10^{19}$~eV. This energy slowly decreases because of pair
 production and adiabatic losses during propagation over larger
 distances, but without change in the shape of the distribution.

%%%%%%%%%%%%%%%%%%%%%%%%%%%%%%%%%%%%%%%%%%%%%
%%%%%%%%%%%%%%%%%%%%%%%%%%%%%%%%%%%%%%%%%%%%%
  The lower panel of Fig.~\ref{power} shows the fractional energy
 carried by different particles after propagation in terms of the
 total energy of the protons injected with energy spectrum
 described by Eq.~\ref{spectrum}. The proton curve, which also
 includes neutrons, always dominates. The energy content in protons,
 however, is only about 50\% of that injected for distances above
 120~Mpc. The rest of the injected energy is distributed between the
 electromagnetic component and neutrinos.  Note the difference between
 the photon (and electron) components from photoproduction (long
 dashed line), and from pair production (short dashed line).  While
 the photoproduction component rises very quickly and changes very
 little after 100~Mpc, the pair production component is almost
 proportional to the distance, as most of the injected protons,
 despite the high threshold of 10$^{19}$ eV, have similar pair
 production losses. At distances of 100 (200) Mpc 51\% (43\%) of the
 injected power is carried by nucleons, 31\% (37\%) by the
 electromagnetic component and 18\% (20\%) by neutrinos. The neutrino
 fluxes will remain at the same level during propagation over larger
 distances, and the established energy balance will only slightly
 change as nucleons yield some of their power to the electromagnetic
 component through pair production. Adiabatic losses will, of course,
 affect all components in the same way.

 In addition to distributing a fraction of the energy of the injected
 protons to secondary particles, the propagation changes the
 energy spectrum of protons. The most energetic nucleons lose
 energy fast and are downgraded after a short propagation distance.  
 The number of nucleon with energy above $10^{21}$~eV decreases
 by 10\%, 50\% and 90\% from the injected number of protons after
 only 1, 6, and 20 Mpc. The corresponding distances for nucleons
 of energy above $10^{20}$ eV are 10, 40 and 85 Mpc.  The magnitude
 of these changes emphasizes the importance of detection of very
 high energy particles: for particles of energy above 3$\times$10$^{20}$ eV
 (same as the highest energy event detected by the Fly's Eye~\cite{FEHi})
 these distances are 1, 10 and 30 Mpc. The rapid absorption of the
 highest energy cosmic rays implies that the horizon of the highest
 energy protons is very small, and increases the energetics requirements
 for potential UHECR sources.

%%%%%%%%%%%%%%%%%%%%%%%%%%%%%%%%%%%%%%%%%%%%%%%%%%%%%%%%%%%%%%%%%%%%%%%%%%%%%%%%

\section{Discussion, conclusions and outlook}

 The Monte Carlo propagation of ultra high energy protons in a random
 extragalactic magnetic field has obvious advantages over other approaches
 to calculations of proton propagation in the cosmologically nearby Universe.
 To start with, this approach takes fully into account 
 fluctuations in the positions of proton interactions, and thus also in the
 proton energy losses and production of secondary particle fluxes.
 It also naturally generates the correlations between the proton's arrival
 energy, its time delay, and its angular deviation from the source direction.
 We have also shown that mathematical approaches which use a diffusion
 description of magnetic scattering, although superior in computational
 speed, can lead to significant systematic errors for propagation
 distances smaller then $\sim 100$ Mpc.

 These features of the calculation become extremely valuable
 when applied to specific models of UHECR acceleration, especially
 models that involve a relatively short (compared to light travel time
 and proton time delay) active phase of the source. An extreme example
 for such a model is the GRB model for UHECR acceleration.  However, other
 models involving interacting galaxies or radio galaxies of specific
 morphology could also be affected, especially if embedded in regions
 of high (random) magnetic field.

 At energies that allow protons to photoproduce, namely above
 10$^{20}$ eV, the energy degradation is extremely rapid. This is not
 very surprising because of the very short photoproduction interaction
 length at energies corresponding to the maximum cross section --
 i.e. $\lambda_{\rm ph}$ below 4 Mpc for energies between
 4$\times$10$^{20}$ eV and 10$^{21}$ eV. This energy range is very relevant,
 as it is just above the
 highest energy particles detected by the Fly's Eye and AGASA
 arrays~\cite{FEHi,AGASA}. A large part of this rapid energy
 dissipation in our calculation is due to the correct implementation
 of the fluctuations in photoproduction interactions in SOPHIA. A good
 example for the size of the fluctuations is the proton energy
 distribution after propagation over 10 Mpc shown in Fig.~\ref{power},
 which covers more than one and a half orders of magnitude. This is an
 extreme case.  However, every particle injected with an energy well above
 the photoproduction threshold would very rapidly result in a distribution
 extending down to the threshold, within the first 10 Mpc.

  This rapid energy dissipation creates additional problems for models
 of cosmic ray acceleration at astrophysical shocks. Apart from the difficult
 question of the maximum acceleration energy, such models require that
 a significant fraction (0.01 to 0.1) of their source luminosity contributes
 to the UHECR flux. The rapid energy dissipation increases the energy
 requirements in terms of total luminosity and severely
 limits the source distance. Because of magnetic scattering, such limits
 could also be set for particles injected with energy below the
 photoproduction threshold.

%%%%%%%%%%%%%%%%%%%%%%%%%%%%%%%%%%%%%%%%%%%%%%%%%%%%%%%%%%%
\begin{figure}[tb] % fig 9
\centerline{\epsfig{file=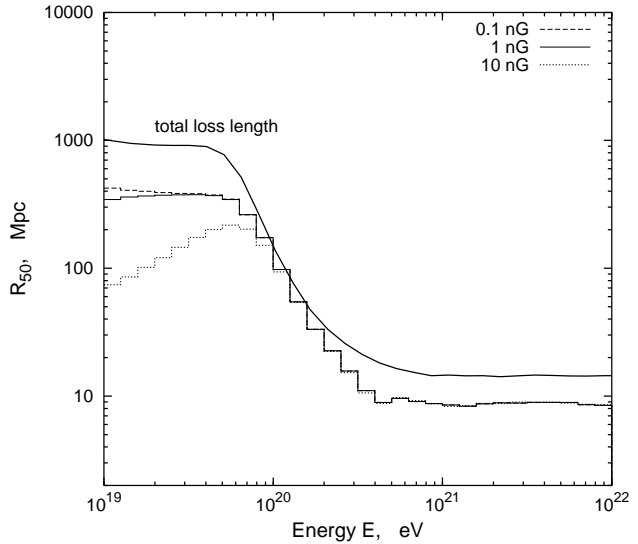,width=85mm}}
\vspace*{10pt}
\caption{Proton 50\% horizon as a function of injection energy for
 average random magnetic fields of 0.1 (dashed histogram), 
1 (solid histogram), and 10 (dotted histogram) $n$G. 
See text for definition. The solid line
is the total energy loss length from Fig.~\protect\ref{fig1}, shown here
for comparison.
}
\label{horiz}
\end{figure}
%%%%%%%%%%%%%%%%%%%%%%%%%%%%%%%%%%%%%%%%%%%%%%%%%%%%%%%%%%%
 Fig.~\ref{horiz} shows the 50\% horizon for UHECR sources as a
 function of source particle energy for $\langle B \rangle$
 values of 0.1, 1 and 10 $n$G. The 50\% horizon $R_{50}$ is defined
 here as the light propagation distance to the source
at which 1/$e$ of all injected 
protons have retained 50\% or more of their energy, 
i.e.\ $R_{50}$ is achieved when
\begin{equation}
  \int_{\frac{E_0}{2}}^{E_0} {d N \over d E} dE = N_0 \exp(-1),
\end{equation}
 where $N_0$ is the number of particles injected with energy $E_0$.

 To start with, $R_{50}$ is small at any energy, and demonstrates
 the resonant nature of the photoproduction cross section.
 At E = 10$^{20}$ eV $R_{50}$ is about 100 Mpc, while at 2$\times$10$^{20}$ eV
 it decreases to 20 Mpc and becomes smaller than 10 Mpc for energies
 above 3$\times$10$^{20}$ eV.  For injection energies  above 10$^{20}$ eV
 the horizon energy dependence is similar to that of the  energy loss
 distance shown in Fig.~\ref{fig1}. These protons are not affected much
 by the magnetic field since their scattering angles are small, but
 suffer mainly from energy degradation due to $p\gamma$ encounters.
 Below 10$^{20}$ eV the picture changes. The scattering in the magnetic
 field increases the propagation time and thus causes additional energy
 loss and an increase of the ratio $x_{\rm loss}/R_{50}$.

 Stronger magnetic fields create delays, that could be longer than
 the light propagation time from the source and reverse the trend --
 the horizon starts decreasing below $\sim 6\times 10^{19}$~eV
 and is restricted to 75 Mpc at 10$^{19}$ eV. Since the average time 
 delay is inversely proportional to $E^2$, the decrease of $R_{50}$
 is expected to become more drastic at lower energy. 
 One consequence of the strong energy dependence of $R_{50}$ is,
 for example, that our attempts to correlate the arrival directions
 of UHECR with different types of astrophysical objects should use only
 objects within the particle horizon depending on the
 magnetic fields strength in different regions of the Universe.
 Independently of the magnetic field value, however, the horizon
 defined above is much smaller than the conventional numbers of
 50 or 100~Mpc for the highest energy cosmic ray events.

 There are many relevant astrophysical problems which can be studied
 with the approach described in this paper.  We plan to use
 the code for proton propagation in regular magnetic fields associated
 with large scale structures (local supercluster, supergalactic plane).
 The regular fields, especially if they reach the observationally
 allowed limits of 0.03~$\mu$G and even 0.1~$\mu$G, could change the propagation
 patterns for 10$^{19}$~eV cosmic ray protons and alter the horizon
 values shown in Fig.~\ref{horiz}. We also plan to set limits on models
 of slow UHECR acceleration on shocks of very large dimensions and
 to look for possibilities of ultra-high energy $\gamma$--ray halos
 around the sources
 and along the tracks of the UHECR protons.

%%%%%%%%%%%%%%%%%%%%%%%%%%%%%%%%%%%%%%%%%%%%%%%%%%%%%%%%%%%%%%%%%%%%%%%%%%%%%%%%

\section*{Acknowledgments}
 The authors are indebted to A.~Achterberg and the authors of
 Ref.~\cite{Acht99} for sharing their corrected results 
 prior to publication, and to P.P.~Kronberg for careful reading
 of the manuscript and valuable discussions.
 The research of TS is supported in part by NASA Grant NAG5-7009.
 RE is supported in part by the US Department of Energy contract
 DE-FG02 91ER 40626. The work of RJP and AM was supported in part
 by a grant from the  Australian Research Council.
 JPR is supported through the TMR network Astro--Plasma Physics,
 funded by the EU under contract FMRX-CT98-0168. 
 AM thanks BRI for its hospitality during her visit. These
 calculations are performed on DEC Alpha and Beowulf clusters funded
 by NSF grant PHY--9601834.
%%%%\ifonecol\raggedright\fi
\bibliographystyle{prsty}
%\bibliography{uhecr,physics}

\begin{appendix}

\section{Monte Carlo sampling of interaction points}

In the following we discuss the application of the veto algorithm to the
sampling of interaction points along a nucleon propagation path.
The probability of having no hadronic interaction with a photon of the
CMB within a path length interval $(s_1,s_2)$ reads
\begin{equation}
P_{\rm no}(s_1,s_2)
 = \exp\left\{ - \int_{s_1}^{s_2} \frac{ds}{\lambda_{\rm ph}(E(s))}
\right\}.
\end{equation}
The interaction length itself depends only on the nucleon energy. 
However, because of the treatment of Bethe-Heitler losses as 
continuous process, this energy depends on the path length $s$.
Correspondingly, the probability for one interaction in the 
interval $(s,  s+ds)$ is given by
\begin{equation}
P_{\rm int}(s) ds = P_{\rm no}(0,s)\ \frac{ds}{\lambda_{\rm ph}(E(s))},
\label{int-point-generic}
\end{equation}
where $P_{\rm no}(0,s)$ is the probability that no interaction has occurred
before. In our approach we replace $\lambda_{\rm ph}(E(s))$ by the
constant $\lambda_{\rm ph,min}$ and use (\ref{int-point-generic}) to
sample the path length distance from the current
location ($s = 0$) to the next interaction. 
This interaction point is then accepted with the
probability $\lambda_{\rm ph,min}/\lambda_{\rm ph}(E(s))$.
Hence the interaction probability can be written as
\begin{eqnarray}
P_{\rm int}(s) ds  &=& 
\Bigg[ \tilde{P}_{\rm no}(0,s)
\nonumber\\
& &\hspace*{-1.5cm} 
+ \int_0^s \frac{ds_1}{\lambda_{\rm ph,min}}
\tilde{P}_{\rm no}(0,s_1) 
\left(1-\frac{\lambda_{\rm ph,min}}{\lambda_{\rm ph}(E(s_1))}\right) 
\tilde{P}_{\rm no}(s_1,s) 
\nonumber\\
& &\hspace*{-1.5cm}
+ \int_{0}^{s} \frac{ds_1}{\lambda_{\rm ph,min}}
\tilde{P}_{\rm no}(0,s_1)
\left(1-\frac{\lambda_{\rm ph,min}}{\lambda_{\rm ph}(E(s_1))}\right)
\nonumber\\
& &\hspace*{-1.5cm}
\times \int_{s_1}^{s} \frac{ds_2}{\lambda_{\rm ph,min}}
\tilde{P}_{\rm no}(s_1,s_2)
\left(1-\frac{\lambda_{\rm ph,min}}{\lambda_{\rm ph}(E(s_2))}\right)
\tilde{P}_{\rm no}(s_2,s)
\nonumber\\
& &\hspace*{-1.5cm}
 +\hspace*{1cm} \dots\hspace*{1cm} \Bigg]
\left(\frac{\lambda_{\rm
ph,min}}{\lambda_{\rm ph}(E(s))}\right) \frac{ds}{\lambda_{\rm ph,min}},
\label{our-method}
\end{eqnarray}
where we have used
\begin{equation}
\tilde{P}_{\rm no}(s_2,s_1) = \exp\left\{
-\frac{s_1-s_2}{\lambda_{\rm ph,min}}\right\}\ .
\end{equation}
The first term in square brackets corresponds to the probability that no
interaction was sampled in the interval $(0,s)$. The second term is the
contribution which comes from an interaction point sampled at $s_1$ but
rejected with the probability $1-\lambda_{\rm ph,min}/\lambda_{\rm ph}$.

The integration limits in (\ref{our-method}) ensure the ordering of the
interaction points according to the simulation method, $0<s_1<s_2<\dots<s$.
Symmetrizing the integration limits yields
\begin{eqnarray}
P_{\rm int}(s) ds &=& \frac{ds}{\lambda_{\rm ph}(E(s))} 
\exp\left\{ -\frac{s}{\lambda_{\rm ph,min}}\right\}
\nonumber\\
& &\hspace*{-1cm}\times\ \sum_{n=0}^\infty \frac{1}{n!} \left[ \int_0^s
ds^\prime
\left(\frac{1}{\lambda_{\rm ph,min}}-\frac{1}{\lambda_{\rm
ph}(E(s^\prime))}\right)\right]^n
\nonumber\\
&=& \exp\left\{ -\int_0^{s}
\frac{ds^\prime}{\lambda_{\rm ph}(E(s^\prime))}
\right\}\ \frac{ds}{\lambda_{\rm ph}(E(s))} ,
\end{eqnarray}
which is identical to (\ref{int-point-generic}) and shows that the
described simulation method reproduces the correct, energy-dependent 
interaction length.

\section{Implementation of the magnetic field}

A turbulent magnetic field which is frozen into a fluid with fully developed
hydrodynamic turbulence would follow a Kolmogorov spectrum, which is defined
by
\begin{equation}
I(k) = I_0 (k/k_0)^{-5/3}
\end{equation}
 where $k$ is the wavenumber~\cite{Kolmogorov}. $I(k)$ is the energy
 density per unit wave number, $k_0$ the smallest wavenumber of the
 turbulence, the inverse $k_0^{-1}$ is sometimes called the ``cell size''
 of the turbulence. Hence we have for the total energy density \cite{BS87}
\begin{equation}
U_{\rm tot} = \frac{B_{\rm rms}^2}{8 \pi} = \int {\rm d}k\,I(k)\ .
\end{equation}
In the propagation program we consider 3 discrete wave numbers.  Thus we
have to rewrite this integral in terms of a discrete spectrum in $k$,
starting with $k_0$ and continuing with $k_i = 2k_{i-1}$, $i=1,2$. These are
equally spaced apart in $\log_2 k$, with $\Delta (\log_2 k)$=1. Hence the
energy density we should ascribe to each of the three wavenumbers is
approximately
\begin{eqnarray}
U_i &\approx& \left. {I(k)\,{\rm d}k \over {\rm d}(\log_2 k)}\right|_{k_i}
\Delta (\log_2 k)\nonumber\\ 
&=& I_0 k_0 \ln 2 \left({k_i \over
k_0}\right)^{-2/3}\ . 
\end {eqnarray}
The total energy density is then a simple sum, 
\begin{equation}
\frac{B^2}{8 \pi} = U_0+U_1+U_2\ .
\end{equation}
We normalize the field to a total energy density corresponding
to $\langle |B| \rangle = 1\,{\rm nG}$, i.e. $U_{\rm tot} \approx
4{\times}10^{-20}\,{\rm erg\,cm}^{-3}$.

The technical implementation of the magnetic field into our propagation code
is as follows. We divide the propagation volume into cubes of $1\,{\rm Mpc}$
side length, and attach to each of them a homogeneous field $\mathbf{B_0}$
with magnitude $B_0$ and random direction. Each of these cubes is divided
into 8 cubes of $0.5\,{\rm Mpc}$ side length, to which a field $\mathbf{B_1}$
of magnitude $B_1$ and random direction is vectorially added to the field
${\mathbf B_0}$. The procedure is repeated once more, so that our field is eventually
realized on elementary cubes of $0.25{\rm Mpc}$ side length, each of which
carries a magnetic field $\mathbf{B_0+B_1+B_2}$. We check that ${\rm div}
{\mathbf B} \simeq 0$ by approximating the surface integral with
 the sum of the outward normal component of ${\mathbf B}$ over the
 surface of the 8$\times$8$\times$128 Mpc$^3$ volume $V$. The volume
 averaged value of div(${\mathbf B}$) is calculated as
\begin{equation}
 \langle {\mathbf \nabla \cdot B} \rangle \; = \;
 {{1} \over {V}} \sum {\mathbf B}_\perp ds \; .
\end{equation}
 The r.m.s. value of $\langle {\mathbf \nabla \cdot B} \rangle$ for 10,000
 field realizations is  $\langle {\mathbf \nabla \cdot B} \rangle_{\rm rms}$
 = $3.7 \times 10 ^{-6}$ $n$G/kpc.

 We also calculate the effective correlation length $\ell_{\rm corr}$ 
 by equating
\begin{equation}
 \langle {\mathbf B}({\mathbf x}) \cdot {\mathbf B}({\mathbf x} +
 {\mathbf \xi}) \rangle
 \; = \; {\mathbf B}^2({\mathbf x}) \exp \left(- \frac{|{\mathbf \xi}|}
 {\ell_{\rm corr}} \right)\ .
\end{equation}
 The best fit value of $\ell_{\rm corr}$  is 390 kpc.
\end{appendix}
\end{document}